\documentclass[jgr,draft]{agujournal2019}
\usepackage{url} 
\usepackage[inline]{trackchanges} 
\usepackage{soul}
\usepackage{multirow}
\usepackage{graphicx}
\usepackage{colortbl}
\usepackage{color}
\usepackage{amssymb}
\usepackage{amsmath}
\usepackage{caption}
\usepackage[normalem]{ulem}

\draftfalse

\journalname{JGR: Space Physics}

\begin{document}

%
%


\title{Multi-scale observation of magnetotail reconnection onset: 1. macroscopic dynamics}

%
%




\authors{	Kevin J. Genestreti\affil{1},
		Charles J. Farrugia\affil{2},
		San Lu\affil{3}\thanks{1},
		Sarah K. Vines\affil{4},
		Patricia H. Reiff\affil{5},
	      	Tai Phan\affil{6},
		Daniel N. Baker\affil{7},
		Trevor W. Leonard\affil{7}\thanks{4},
		James L. Burch\affil{8},
		Samuel T. Bingham\affil{4}\thanks{2},
		Ian J. Cohen\affil{4},
		Jason R. Shuster\affil{9}\thanks{3},
		Daniel J. Gershman\affil{9},
		Christopher G. Mouikis\affil{2},
		Anthony J. Rogers\affil{2},
		Roy B. Torbert\affil{1,2},
		Karlheinz J. Trattner\affil{7},
		James M. Webster\affil{5,8},
		Li-Jen Chen\affil{9},
		Barbara L. Giles\affil{9},
		Narges Ahmadi\affil{6},
		Robert E. Ergun\affil{6},
		Christopher T. Russell\affil{3},
		Robert J. Strangeway\affil{3},
		Rumi Nakamura\affil{10}}


\affiliation{1}{Earth Oceans and Space, Southwest Research Institute, Durham, New Hampshire, USA}
\affiliation{2}{Earth Oceans and Space, University of New Hampshire, Durham, New Hampshire, USA}
\affiliation{3}{Institute for Geophysics and Planetary Physics, University of California Los Angeles, Los Angeles, California, USA}
\affiliation{4}{Applied Physics Laboratory, Johns Hopkins University, Laurel, Maryland, USA}
\affiliation{5}{Rice Space Institute, Rice University, Houston, Texas, USA}
\affiliation{6}{Space Science Laboratory, University of California Berkeley, Berkeley, California, USA}
\affiliation{7}{Laboratory for Atmospheric and Space Physics, University of Colorado Boulder, Boulder, Colorado, USA}
\affiliation{8}{Space Science and Engineering Division, Southwest Research Institute, San Antonio, Texas, USA}
\affiliation{9}{Goddard Space Flight Center, National Aeronautics and Space Administration, Greenbelt, Maryland, USA}
\affiliation{10}{Space Research Institute, Austrian Academy of Sciences, Graz, Austria}
\thanks{1}{Now at School of Earth and Space Sciences, University of Science and Technology of China, Hefei, Anhui 230026, China}
\thanks{2}{Deceased}
\thanks{3}{Now at Space Science Center, Institute for the Study of Earth Oceans and Space, University of New Hampshire, Durham, New Hampshire, USA}
\thanks{4}{Now at Cooperative Institute for Research in Environmental Sciences, CU, Boulder, and National Oceanic and Atmospheric Administration National Centers for Environmental Information}




\correspondingauthor{Kevin J. Genestreti}{kevin.genestreti@swri.org}




\begin{keypoints}
\item Magnetotail reconnection onset was observed during a fortuitous multi-scale conjunction of the heliophysics observatories
\item A transient solar wind pressure pulse triggered thinning and stretching of the cross-tail current sheet without significant flux loading
\item A second solar wind pressure pulse caused the thinned current sheet to rapidly collapse and reconnect
\end{keypoints}

%
%

%
%


\begin{abstract}
We analyze a magnetotail reconnection onset event on 3 July 2017 that was observed under otherwise quiescent magnetospheric conditions by a fortuitous conjunction of six space and ground-based observatories. The study investigates the large-scale coupling of the solar wind - magnetosphere system that precipitated the onset of the magnetotail reconnection, focusing on the processes that thinned and stretched the cross-tail current layer in the absence of significant flux loading during a two-hour-long preconditioning phase. It is demonstrated with data in the (1) upstream solar wind, (2) at the low-latitude magnetopause, (3) in the high-latitude polar cap, and (4) in the magnetotail that the typical picture of solar wind-driven current sheet thinning via flux loading does not appear relevant for this particular event. We find that the current sheet thinning was, instead, initiated by a transient solar wind pressure pulse and that the current sheet thinning continued even as the magnetotail and solar wind pressures decreased. We suggest that field line curvature induced scattering (observed by Magnetospheric Multiscale (MMS)) and precipitation (observed by Defense Meteorological Satellite Program (DMSP)) of high-energy thermal protons may have evacuated plasma sheet thermal energy, which may require a thinning of the plasma sheet to preserve pressure equilibrium with the solar wind. 
\end{abstract}

%
%

%


%
%
%
%

\section{Introduction}

Magnetospheric substorms are triggered by magnetotail reconnection, which, in turn, follows thinning and stretching of the cross-tail current sheet \cite{Hones.1979}. The commonly-accepted processes that thin and stretch the current sheet require solar wind driving of the magnetosphere with southward interplanetary magnetic fields (IMF) $B_Z\lesssim0$ \cite{Baker.1996}. Magnetic reconnection between the IMF and low-latitude dayside magnetopause ``opens'' magnetospheric field lines that are then convected over the geomagnetic poles into the high-latitude magnetotail lobes \cite{Dungey.1961}. Magnetic flux loading causes the tail lobes to flare outward at a larger angle, intruding further into the solar wind ram flow \cite{McPherron.2002}. For steady solar wind dynamic pressure the tail pressure must increase to maintain equilibrium, which results in compression of the equatorial tail current sheet. Simultaneously, large-scale pressure gradients drive a return flow of high-entropy flux tubes from the near-Earth equatorial tail back to the magnetopause that is eroded by dayside reconnection \cite{HsiehandOtto.2015}. Again, the thickness of the equatorial current layer is reduced to maintain vertical pressure balance. Magnetotail reconnection is common during northward IMF too \cite{Zhang.2016}, and the mechanisms that drive current sheet thinning and stretching without low-latitude dayside reconnection are not understood.

Isotropic proton precipitation in the high-latitude ionosphere is a well-known symptom of cross-tail current sheet thinning \cite{Sergeev.1983,Donovan.2012}. As the tail current sheet thins down to the proton-kinetic scale, non-adiabatic scattering drives pitch-angle diffusion in previously trapped current sheet protons. Protons that become sufficiently field aligned stream out of the current sheet and into ionosphere and neutral atmosphere, where they are effectively lost from the tail. This pitch-angle scattering is most efficient when particles' gyroradii $R_G(E_\bot)$ are comparable to the magnetic field line curvature (FLC) radii $R_C$ in the central current sheet; theoretical works consider a critical range of $1\lesssim\kappa(=\sqrt{R_C/R_G})\lesssim3$ \cite{Sergeev.1983,BuchnerandZelenyi.1987,BuchnerandZelenyi.1989,Delcourt.1996}. Thus, FLC pitch-angle scattering acts like a bandpass filter that ejects protons most efficiently over a range of perpendicular energies $E_\bot$, and as a thick ($R_C>>R_G$) current sheet thins, FLC scattering acts most efficiently on lower and lower energy protons. Field line entropy $pV^{5/3}$ (where $p$ is the pressure and $V$ is the volume of a flux tube element, as in \citeA{Birn.2009} equation 4) is typically assumed to be conserved without reconnection \cite{Birn.2009} and, since FLC scattering will reduce the plasma sheet thermal energy in the absence of refilling, lossy FLC scattering and current sheet deformation and/or thinning may go hand-in-hand.

\begin{figure}
\noindent\includegraphics[width=39pc]{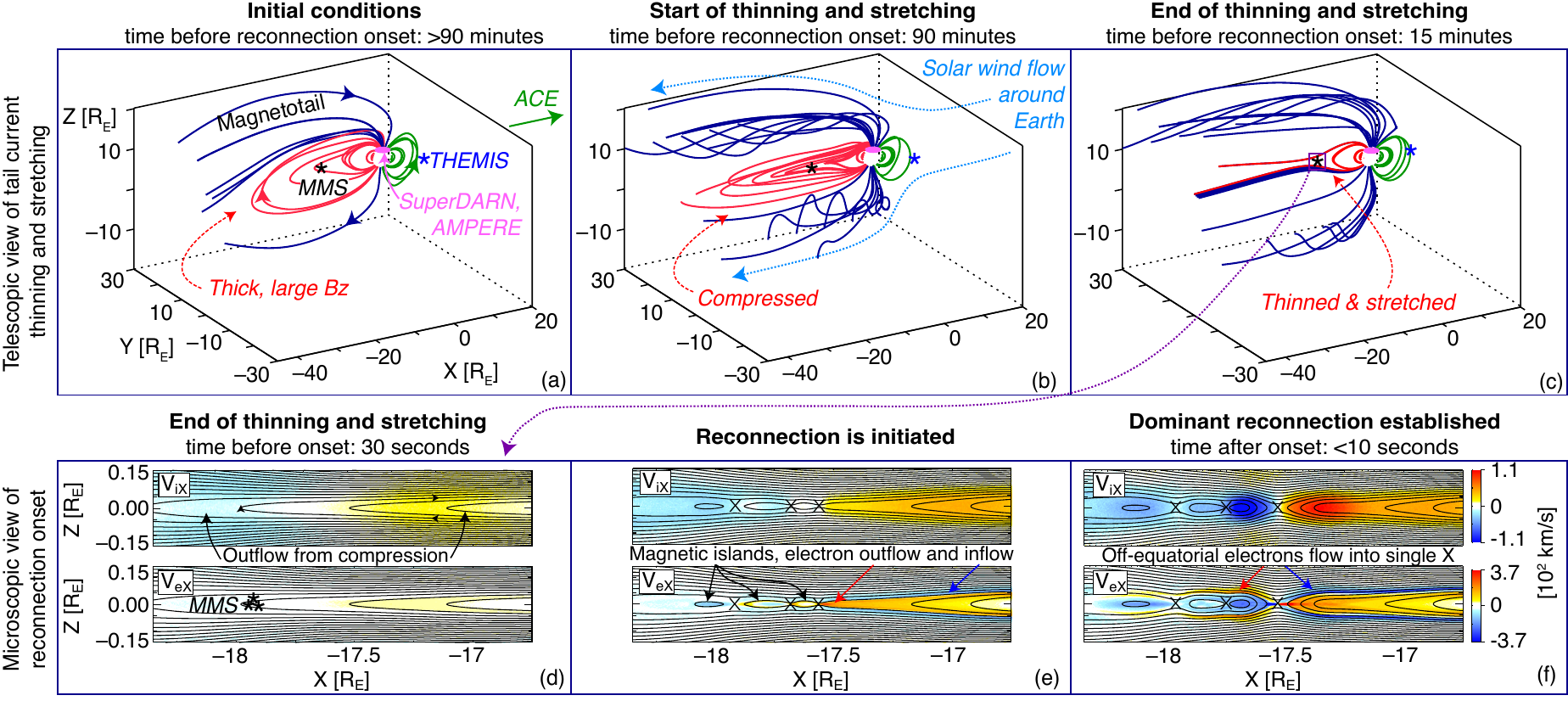}
\caption{(Top) large-scale and (bottom) small-scale physics of magnetotail reconnection onset, illustrated by (top) a global MHD and (bottom) a fully-kinetic 2.5-D PIC simulation. (a) and (d) illustrate the coverage of our multi-scale constellation of observatories. This paper focuses on the macro-scale dynamics while the companion focuses on the micro-scale physics. The axes in (a)-(c) correspond to Geocentric Solar Magnetic (GSM) coordinates, while those in (d)-(f) are arbitrary coordinates locally normal and tangential to the tail current sheet. Times in (a)-(c) are relative to the reconnection onset time observed by MMS at 5:23 UT. Times in (d)-(f) are relative to the onset of reconnection in the PIC simulation.}
\label{modeling}
\end{figure}

In this study, we analyze a fortuitous conjunction of six spacecraft and ground-based observatories, which occurred on 3 July 2017, to determine the global causes of one magnetotail reconnection event. An overview of the regions covered by spacecraft and ground-based observatories is provided in figure \ref{modeling}a, and a summary of the key observations follows. The magnetotail current sheet was initially thick, more dipolar, and stable (Section 3.1). The IMF $B_Z$ was weak or strongly positive and low-latitude dayside reconnection was expected to be weak or fully disabled based on the atypically large plasma $\beta$ observed in the magnetosheath and low magnetic shear at the magnetopause (Section 3.2). Observed patterns in the global field-aligned current system indicated a preference for magnetopause reconnection poleward of the cusps (often referred to as high-latitude reconnection), which does not supply energy to the high-latitude tail or erode the low-latitude dayside. Ionospheric observations showed weak convection in the polar cap that was, at times, either weakly anti-sunward (indicative of weak magnetotail loading by low-latitude dayside reconnection) or sunward (indicating the predominance of reconnection poleward of the magnetopause cusps). The polar cap boundary moved equator-ward by $\sim1^\circ$ prior to reconnection onset, which is a signature of open-flux loading (Section 3.3). These data are somewhat consistent with a global magnetospheric simulation of the event, which predicts weak ionospheric convection but a larger $\sim5^\circ$ expansion of the polar cap (Section 3.4). Despite the lack of significant evidence for solar wind driving, the magnetotail current sheet thinned and stretched substantially (Section 3.5) until the impact of a (second) solar wind pressure pulse precipitated the collapse of the current sheet down to electron scales and reconnection onset (Section 3.6). We note that the tail pressure decreased during the thinning phase and magnetic flux erosion. We conclude that the thinning may have been driven by proton FLC scattering, which was observed in situ in the plasma sheet and also at the ionospheric foot point. A companion paper (hereafter Paper 2) focuses on the microphysics observed by MMS during the reconnection onset.

This paper is organized as follows: in the next section we provide brief overviews of our data set, analysis techniques, empirical models, and physics-based simulation. For brevity, we provide citations with more rigorous descriptions of the instrumentation, data sets, and models. In section 3 we analyze the data and compare with our simulation. Finally, in section 4, we discuss the implications of our findings within the context of the standard picture of substorms. We stress the need for future modeling and data collection endeavors to confirm or refute the findings that were made here, which interpret the multi-scale physics of our sparsely covered magnetosphere. 

\section{Data, analysis methods, and models}

\subsection{Data and analysis methods} 

The Advanced Composition Explorer (ACE) spacecraft orbits the Earth-sun L1 point and provides in-situ measurements of the upstream solar wind plasma. Data are obtained from the OMNI database and are (1) propagated to Earth's bow shock and (2) available at a cadence of 1-per-min. 

The Time History of Events and Macroscopic Interactions during Substorms (THEMIS) mission \cite{Angelopoulos.2008a} provides our study with two-point in-situ measurements of the low-latitude, dusk-side magnetopause. Plasma and magnetic field data from THEMIS satellites D and E are used to (1) obtain the plasma conditions in the shocked solar wind upstream of the magnetopause, (2) identify THEMIS magnetopause crossings, and (3) monitor for signatures of low-latitude reconnection, specifically Alfv\'enic plasma jets at the magnetopause. THEMIS-D, which was located at $\left[XYZ\right]_{GSM}=[6.8, 8.7, -2.6]$ $R_E$ on 3 July 2017 at 4:00 UT, was located in the magnetosheath for most of the event. THEMIS-E, which was closer to Earth than THEMIS-D ($\left[XYZ\right]_{GSM}=[6.8, 3.6, -3.3]$ $R_E$ on 3 July 2017 at 4:00 UT), was located in the magnetosphere during most of the event. However, both spacecraft fully and/or partially crossed the magnetopause multiple times. For THEMIS-D, magnetopause crossings occurred near the tail reconnection onset time ($\sim$5:20 UT) on the inbound leg of its orbit. For THEMIS-E, magnetopause crossings occurred after the impacts of the solar wind pressure pulses, which compressed the magnetopause down to the spacecraft location. Plasma particle fluid moments and omnidirectional fluxes are available once every 4.2 seconds. Magnetic field vector measurements are available once every 62.5 milliseconds. When necessary, simple boxcar averaging is used to downsample the magnetic field to the plasma particle data cadence.

The Active Magnetosphere and Planetary Electrodynamics Response (AMPERE) mission is a constellation of 66 low-altitude satellites, which provides this study with a ``remote'' view of the global magnetospheric current system by measuring, in situ, multi-point magnetic perturbations. These AMPERE magnetic field data are used to derive the global field-aligned current (FAC) configuration at high northern latitudes. FACs indicate magnetospheric magnetic stresses, which can be generated by, among other processes, magnetic reconnection at the magnetopause and magnetotail. Snapshots of the FACs are accumulated over the course of 10 minutes, are available at 1 snapshot per 10 min, and are accumulated by no fewer than 11 spacecraft. The sensitivity of the magnetometers are such that currents below 0.075 $\mu$A/m$^2$ are understood to be indiscernible from noise. Currents below this threshold are discarded in this study. 

The Super Dual Auroral Radar Network (SuperDARN) is a network of ground-based high-frequency radar stations, which provides this study with a remote view of the northern polar cap area and convection electric field, which are derived from maps of plasma convection. The motion of the polar cap boundary indicates whether opened magnetic field lines are being accumulated (polar cap area growth) or closed (polar cap area reduction). The Heppner-Maynard convection boundary \cite{HeppnerandMaynard.1987}, a proxy for the open-closed field line boundary, is located at the lowest latitude in which no fewer than three radars measure a line-of-sight velocity of 100 m/s along the zero-potential contour of the polar cap electric field \cite{Imber.2013}. (Note: the convection boundary could not be derived reliably from AMPERE data as the field-aligned currents were too weak to provide robust fits.) The cross-polar-cap potential is the total potential drop across the polar cap area, and is used as an indicator of the amount of polar ionospheric convection. 

Magnetospheric Multiscale (MMS) is a constellation of four very-closely-spaced satellites that provide the study with in-situ, high-spatiotemporal-resolution measurements of the plasma particles and electric and magnetic fields \cite{Burch.2016a}. During this event, on 3 July 2017 at 5:20 UT, MMS was at $X_{GSM}$=$-$17 $R_E$ and $Y_{GSM}$=3 $R_E$. MMS-1, 2, and 3, separated by 26 km, effectively provide this study with a point measurement of the larger magnetotail dynamics. For this event, no data was available from MMS-4. When unspecified, plots or analyses of ``MMS data'' use averaged data from these three satellites. The plasma ion and electron fluid moments from the fast plasma investigation (FPI) sensors \cite{Pollock.2016} are used to determine the cross-tail current density as $J_y=en_i(v_{iy}-v_{ey})$. The magnetic field, FPI ion and electron data, and energetic ion spectrometer \cite{Mauk.2016} data are used to find the total magnetohydrodynamic pressure. Magnetic field data from the fluxgate magnetometers are determined to within a precision of $\leq$0.1 nT per component \cite{Torbert.2016a}. Magnetic field and plasma fluid moments are used to determine the time-dependent half-thickness of the cross-tail current sheet via the Harris approximation, $h(t)=(B_0^2(t)-B_x^2(t))/\mu_0B_0(t)J_y(t)$ \cite{Thompson.2005}. Here, $B_0$ is the magnetic field strength in the plasma sheet boundary layer, approximated as 60\% of the lobe magnetic field strength or $B_0=0.6B_{lobe}=0.6\sqrt{B^2+\mu_0P_{therm}}$, where $P_{therm}$ is the particle thermal pressure. The fraction $B_0/B_{lobe}=0.6$ was chosen to match observations of the plasma sheet boundary when it is first observed during interval of current sheet flapping at 3 July 2017 5:23 UT. The 60\% is slightly larger than the typical range of $0.3\lesssim B_0/B_{lobe}\lesssim 0.5$ \cite{Petrukovich.2015}. For reference, we also calculate the half-thickness using $B_0/B_{lobe}=1$, as in \citeA{Thompson.2005}, which is relevant when a thin current sheet is not embedded within the thicker plasma sheet. Prior to calculating $h$, long (10-min) time averaging is performed to reduce noise in the initially very weak $J_y$. Ion fluxes from FPI were combined with proton fluxes from the energetic ion spectrometer to determine the scalar ion thermal pressure and total MHD pressure (figs \ref{s}n and \ref{precond}f-g). Whenever necessary, the magnetic field data are resampled at the plasma particle cadence via boxcar averaging. In this study, survey-rate data are used, meaning the plasma particle and magnetic field data are obtained at cadences of once per 4.5 seconds and 8 per second, respectively.

Defense Meteorological Satellite Program (DMSP) is a constellation of low-altitude satellites that collects space weather data, including ions and electron fluxes at approximately 850 km altitude. For the event studied here, two passes of DMSP F-16 came very near the modeled footprint of MMS. DMSP F-16 particle flux data are used to qualitatively evaluate whether plasma sheet particles near MMS may be lost to the ionosphere.

The Synchronous Orbit Particle Analyzer (SOPA) instruments onboard the geosynchronous-orbiting Los Alamos Nation Laboratory GEO satellites measure energetic ion (50 keV to 50 MeV) and electron (50 keV to $\geq$1.5 MeV) fluxes. These data were used to look for particle injections into the near-midnight inner magnetosphere driven by the Earthward-propagating jets from magnetotail reconnection. The methodology used was straightforward; data taken near midnight were scanned at/near the time MMS observed reconnection signatures. Dispersed or dispersionless injections are identified as gradual or steep enhancements in the fluxes of energetic particles in energy-time-flux diagrams \cite{Sarris.1976}. No such signatures were identified, indicating that the Earthward jets from the reconnection did not penetrate to geosynchronous orbit. No SOPA data are shown.

\subsection{Empirical models and indices}

We use the T96 empirical model to trace magnetospheric field lines \cite{Tsyganenko.1995,Tsyganenko.1996}. Field-line tracing is used to (1) estimate the outer-magnetospheric source region of the field-aligned currents observed by AMPERE and (2) estimate the proximity between DMSP and the ionospheric foot points of magnetotail field lines near MMS. 

We use the empirical maximum magnetic shear model \cite{Trattner.2007} to estimate the location of the magnetopause reconnection region relative to THEMIS and estimate the outflow direction expected for any THEMIS-observed reconnection outflows. The maximum shear model uses the Cooling empirical model \cite{Cooling.2001} for the IMF draping about the T96 magnetopause. The location of the reconnection line is determined as the line of maximum sheared magnetic energy. The shear model is also used to estimate the magnetic shear angle at the subsolar magnetopause. 

We use the empirically-derived Boyle index \cite{Boyle.1997} to estimate the cross-polar-cap potential from ACE solar wind data, which is found to be in good agreement with SuperDARN-derived values. We use the northern polar cap (PCN) index as a proxy for the strength of convection in the polar ionosphere \cite{Troshichev.1988}. Future work is needed to compare SuperDARN convection maps with the PCN and magnetosphere-ionosphere simulation. We used the $\epsilon$ parameter to approximate the energy input rate (from the solar wind into the magnetosphere via dayside reconnection) per-unit-magnetopause-area, where $\epsilon=V_{SW}B_{SW}^2/\mu_0\mathrm{sin}^4(\theta/2)$ \cite{PerreaultandAkasofu.1978} and $\theta$ is the clock angle of the IMF in the $Y-Z_{GSM}$ plane. 

\subsection{Physics-based simulations}

We use the Space Weather Modeling Framework / Block-Adaptive-Tree-Solarwind-Roe-Upwind-Scheme (SWMF/BATS-R-US) model \cite{Toth.2012} to simulate the global magnetospheric-ionospheric dynamics during 3 July 2017 from midnight to 8:00 UT. The real time-dependent dipole tilt angle is used. Upstream solar wind conditions are obtained from ACE via the OMNI database.  The high-resolution grid with $>$9.6 million cells was used. The modeled ionospheric conductance was determined self-consistently by geomagnetic field-aligned currents. The model is used to (1) provide qualitative visualizations of the magnetospheric configuration (e.g., Fig. \ref{modeling}) and (2) compare with low-altitude and ionospheric observations by AMPERE spacecraft and SuperDARN radars. The full inputs and outputs of our model run are publicly available on the Community Coordinated Modeling Center (CCMC).

The particle-in-cell simulation shown in Figure \ref{modeling}d-f was performed using the same fully-kinetic 2.5-d code as in \cite{Lu.2020,Lu.2022}. The initial conditions were chosen to roughly correspond to those observed by MMS near the time of reconnection onset. The initial current sheet had a finite $B_Z$. Reconnection was initiated by applying a brief pulse of the electric field $E_Y$ to the regions upstream of the current sheet early in the run. The pulse in $E_Y$ imitates the solar wind-driven compression of the high-latitude magnetosphere. Selected frames from the run, shown in Figure \ref{modeling}d-e, are used to illustrate the general dynamics of reconnection onset, and no detailed analysis or comparison with MMS is performed.

\section{Data and model analyses}

The event, which occurred on 3 July 2017, is described in terms of three phases, labeled (between Fig. \ref{s}i and \ref{s}j) and demarcated by two vertical dashed lines $\sim$3:19 and $\sim$5:18 UT. The phases are (1) initial conditions, (2) preconditioning, and (3) reconnection onset, which describe the characteristic processes occurring in the magnetotail. The preconditioning phase is characterized by slow current sheet thinning (Fig. \ref{s}l) and stretching (Fig.\ref{s}k).

\subsection{The initially quiet and stable tail}

\begin{figure}
\noindent\includegraphics[width=33pc]{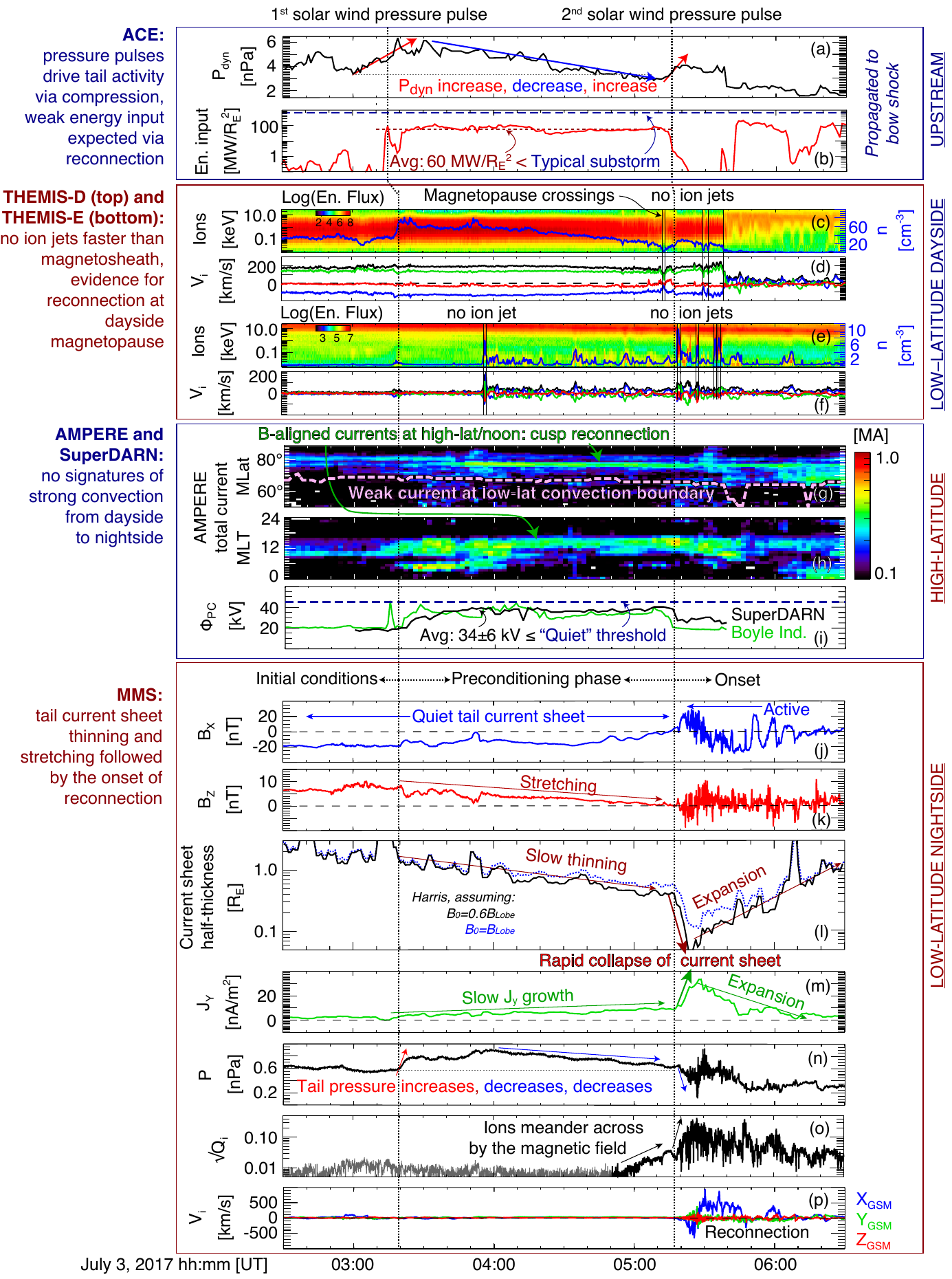}
\caption{(a) solar wind dynamic pressure, (b) derived solar wind energy input, $\epsilon$ (see section 2.2), (c) THEMIS-D ion energy flux (color) and density (blue line), (d) THEMIS-D ion bulk flow velocity, (e) and (f) are the same as (c) and (d) respectively, but for THEMIS-E, (g)-(h) AMPERE total current as a function of MLat and MLT (respectively), integrated over (g) MLT and (h) MLat, where the pink line is the Heppner-Maynard boundary, (i) cross-polar-cap potential from SuperDARN (black) and the Boyle index (green), (j)-(k) MMS-observed $B_X$ and $B_Z$, respectively, (l) the cross-tail current sheet thickness approximated for boundary conditions of 60\% (black) and 100\% (blue) of the lobe magnetic field, (m)-(p) MMS-observed cross-tail current density (m), total MHD pressure (n), ion non-gyrotropy (o), and ion bulk flow vector (p). Two vertical dashed lines delineate the phases of magnetotail activity labeled above (j). The vertical lines jog rightward between panels (b) and (c) to account for small timing differences between the solar wind drivers (a)-(b) and their observed impacts on the magnetosphere (c)-(p).}
\label{s}
\end{figure}

The initial state of the tail is shown in Fig. \ref{s}j-p to the left of the first vertical dashed line, i.e., before 3:20 UT. The orientation of the geocentric solar magnetic (GSM) coordinate axes relative to the magnetosphere are shown in Fig. \ref{modeling}a. The northward component of the equatorial magnetic field ($B_z\approx5-to-10$ nT, Fig. \ref{s}k) magnetizes ions and electrons. The thick current sheet ($h\approx$1-to-2 $R_E$, Fig. \ref{s}l) does not enable the bulk ion population to meander across field lines, as is further evidenced by the negligible ion pressure non-gyrotropy ($\sqrt{Q_i}\lesssim1\%$, Fig. \ref{s}o). Reconnection, which requires slippage of magnetic fields through plasma particles, is neither expected nor observed in this initial tail configuration.

\subsection{Solar wind drivers during preconditioning}

\begin{figure}
\noindent\includegraphics[width=39pc]{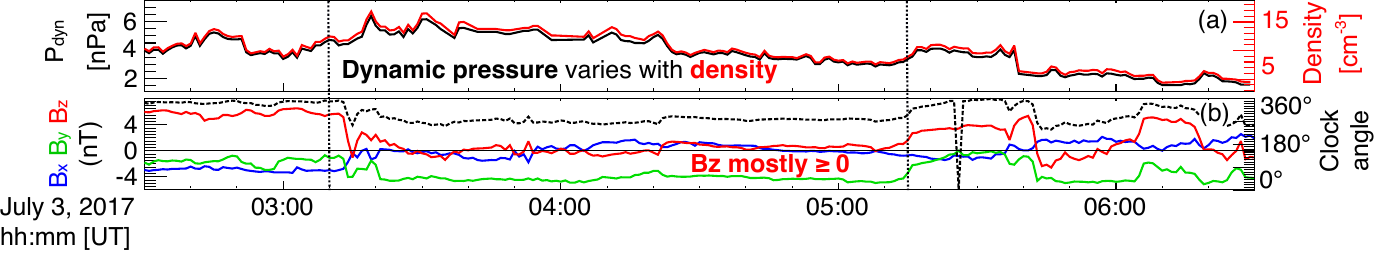}
\caption{(a) Solar wind pressure (black) and density (red), (b) magnetic field vector in GSM (solid) and clock angle (dashed).}
\label{sw}
\end{figure}

The ``inciting incident'' was a transient $\sim$50$\%$ increase in the solar wind dynamic pressure $\Delta P_{dyn,1}$ (first dashed vertical line, Fig. \ref{s}a). $\Delta P_{dyn,1}$ corresponded to a pulsed increase in the solar wind (Fig. \ref{sw}a) and magnetosheath (blue line, Fig. \ref{s}c) densities, and a significant compression of the magnetopause inward to near the location of THEMIS-E at $\left[X,Y,Z\right]_{GSM}$ = [6.8, 3.6, --3.3] $R_E$ (first set of solid vertical lines, Fig. \ref{s}e). After $\Delta P_{dyn,1}$, the solar wind energy input was elevated, but remained lower than the typical rate for a substorm \cite{Akasofu.1981} by roughly an order of magnitude (Fig. \ref{s}b). The rise in the energy input rate corresponded to a rotation of the IMF from mostly northward to mostly duskward (Fig. \ref{sw}b). 

THEMIS-E partially crossed the magnetopause at 3:56 UT, between the first and second pressure pulses ($\Delta P_{dyn,1}$ and $\Delta P_{dyn,2}$, respectively). No clear reconnection jets were observed during the crossings (first set of vertical lines, Fig. \ref{s}e-f). Strong and bipolar ion flows observed during the partial crossing indicate the fast inward-then-outward motion of the magnetopause during compression. THEMIS-D partially crossed the magnetopause twice at 5:11 and 5:15 UT (vertical lines in Fig. \ref{s}c-d). Dayside reconnection exhausts were not observed during these THEMIS-D crossings. 

\begin{figure}
\noindent\includegraphics[width=30pc]{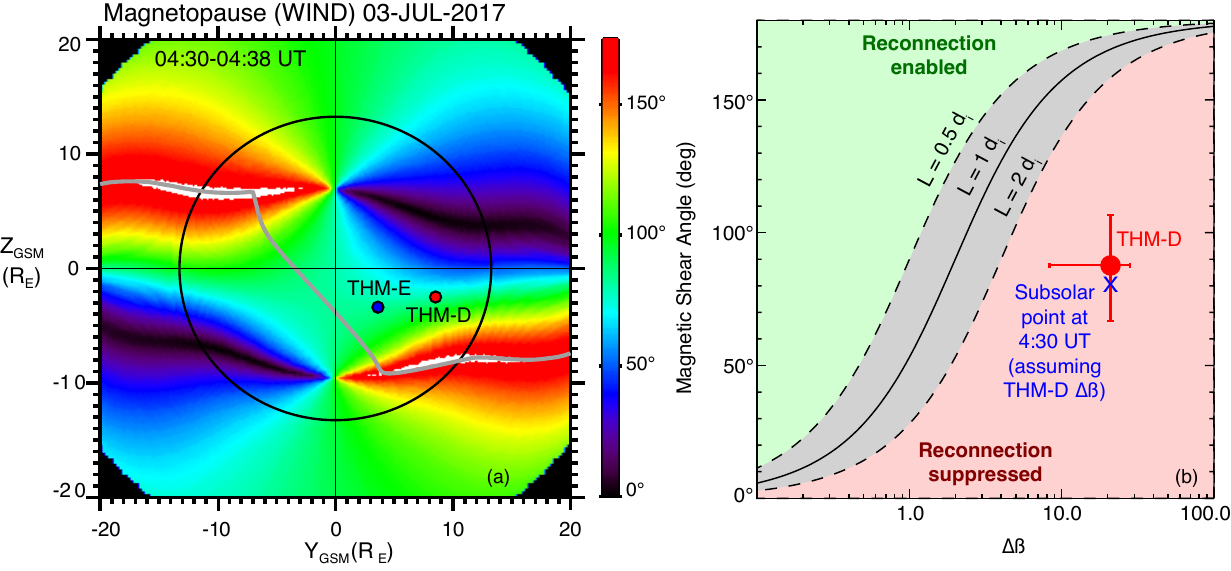}
\caption{(a) The magnetopause magnetic shear angle (color) and X-line location (grey line) predicted by the maximum magnetic shear model, projected into the $Y_{GSM}-Z_{GSM}$ plane. The location of the terminator is marked with a black circle. (b) The average (red circle) and full range (red lines) of $\Delta\beta$ magnetic shear angles observed by THEMIS-D at the magnetopause during preconditioning, where time-varying magnetosheath and average magnetospheric values are used. Magnetospheric $\beta$ and $\vec{B}$ are determined from THEMIS-D at 6:30-6:45 UT. Equivalent values for the subsolar point (blue X) at 4:30 UT are shown, where the modeled shear angle and THEMIS-observed $\Delta\beta$ are used.}
\label{ms}
\end{figure}

Low-latitude reconnection is suppressed by large plasma $\beta$ gradients and low magnetic shear across the magnetopause \cite{Swisdak.2003,Phan.2013}. The suppressed/enabled regimes of reconnection are often depicted as in Fig \ref{ms}b, where reconnection is suppressed (enabled) in the region below (above) a rectifier curve defined by $\Delta\beta=L/d_i\mathrm{tan(\theta/2)}$, where $L$ is the magnetopause thickness and $d_i$ is the ion inertial length. During the preconditioning phase (between the two vertical dashed lines in Fig. \ref{s}e), the expected magnetic shear was roughly 80$^\circ$ at subsolar magnetopause and $\sim$60$^\circ$-to-110$^\circ$ near the THEMIS-D location (Fig. \ref{ms}a). THEMIS-E did not fully cross the magnetopause during the period shown in Fig. \ref{s} and the magnetosheath conditions near THEMIS-E cannot be determined. The plasma $\beta$ observed by THEMIS-D in the high-density magnetosheath was initially $\beta\approx400$ at the time of the solar wind pressure pulse impact following the arrival of the density enhancement in observed by ACE (Fig. \ref{sw}a), and it remained $\beta\geq100$ until roughly 5:00 UT. THEMIS-D remained in the magnetosheath and magnetosphere boundary layer until after 6:00 UT, meaning that $\Delta\beta$ could not be calculated for each partial magnetopause crossing. Once in the magnetosphere, however, THEMIS-D observed $\beta$=0.4, which is used with the time-dependent magnetosheath $\beta$ to calculate $\Delta\beta$ (note that since the magnetospheric $\beta$ is typically small, $\Delta\beta$ is dominated by the magnetosheath $\beta$). The magnetic shear angle and $\Delta\beta$ indicate that low-latitude reconnection was likely suppressed at the THEMIS-D location, and possibly at the subsolar point, in the $\sim$2 hours after the impact of the solar wind pulse (Fig. \ref{ms}b). 

\subsection{Polar cap during preconditioning: data}

During the preconditioning phase (between the vertical dashed lines, Fig. \ref{s}), the strongest field-aligned currents were observed by AMPERE at high magnetic latitudes (MLat; Fig. \ref{s}g) at post-noon magnetic local time (MLT; Fig. \ref{s}h). Field lines in this region are open and map to the high-latitude magnetopause poleward of the cusp. The dusk-ward skew of the FACs is consistent with high-latitude reconnection for strong IMF $B_y<0$ \cite{BurchandReiff.1985}. High-latitude, open-field-line reconnection does not drive day-to-night flux transport, though it enables the penetration of the IMF $B_y$ into the lobes, which can torque and twist the cross-tail current sheet \cite{Crooker.1979,TsyganenkoandSitnov.2007}.

Much weaker currents are observed near the low-latitude Heppner-Maynard convection boundary (Fig. \ref{s}g, dashed pink line). These lower-latitude currents are near the pre-noon open-closed field line boundary, and may indicate weak reconnection of closed field lines. The latitude of the strongest FACs did not move appreciably either poleward or equator-ward after 3:43 UT, which indicates that the magnetopause was not being eroded by closed-field-line reconnection \cite{Coxon.2014}. The northern polar cap index PCN suggests that weak anti-sunward convection occurred during and shortly after $\Delta P_{dyn,1}$, while reverse convection occurred at all other times studied here (Fig. \ref{mhdpc}a). Further analysis of SuperDARN convection maps is needed to verify this picture. Both the PCN and cross-polar-cap potential (see Fig \ref{s}i) were weak, with the latter being within the ``quiescent'' threshold of \citeA{Oliver.1983}. The observed cross-polar-cap potential (black line Fig \ref{s}i) is in excellent agreement with the empirically-modeled approximation (green line Fig \ref{s}i) \cite{Boyle.1997}. 

In summary, convection was very weak during the preconditioning phase and the predominance of open-field-line reconnection indicates that significant day-to-night flux transport is not expected.

\subsection{Polar cap during preconditioning: simulation}

\begin{figure}
\noindent\includegraphics[width=42pc]{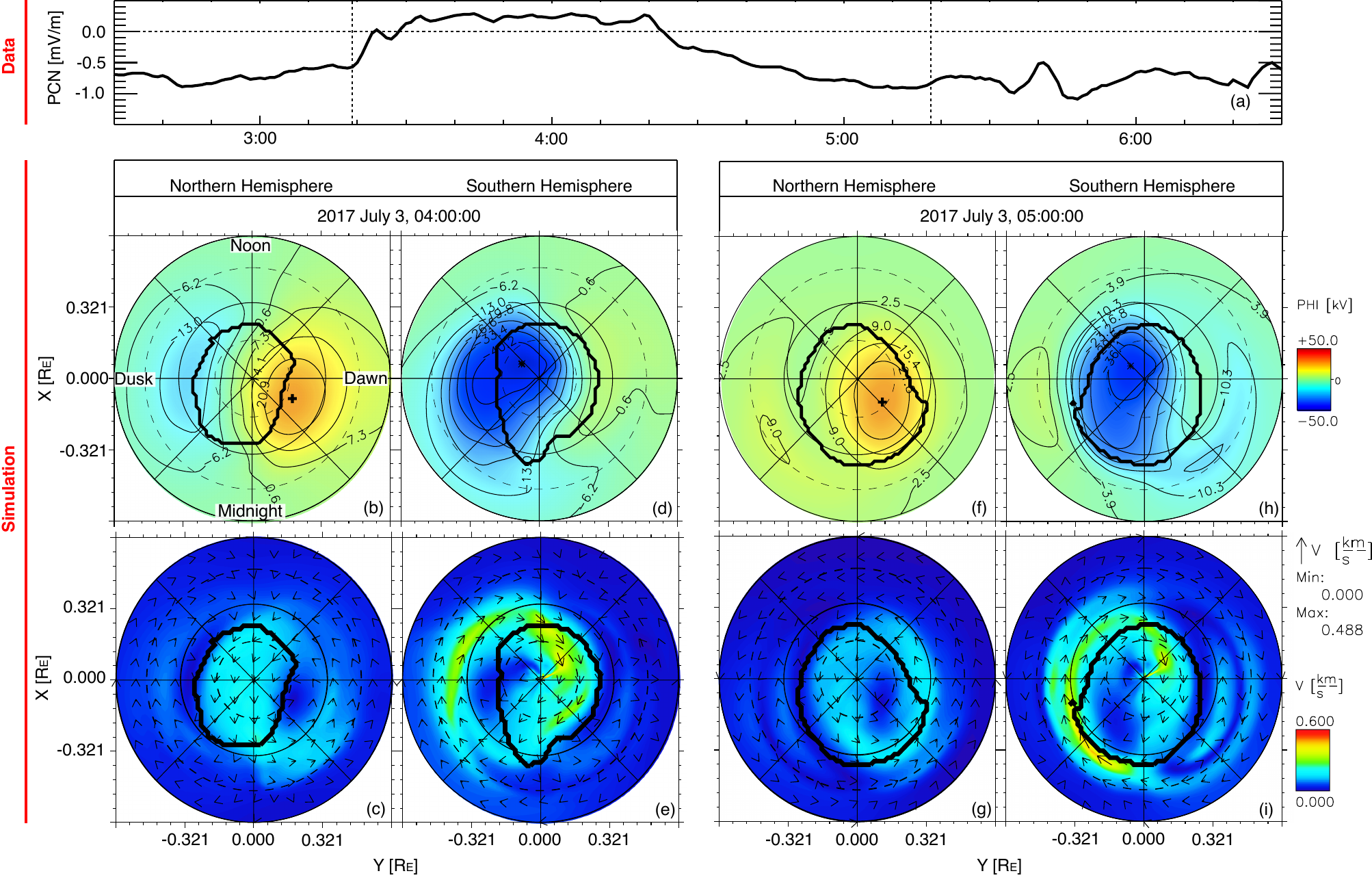}
\caption{(a) The PCN index from data (i.e., not from the simulation). (b-i) Simulated polar cap dynamics during the (b-e) early and (f-i) late stages of the preconditioning phase for the (b-c and f-g) northern and (d-e and h-i) southern polar cap. The top row shows the electric potential in color and contours. The bottom row shows the speed (colors) and velocity vector (arrows). The thick black lines indicate the polar cap boundaries.}
\label{mhdpc}
\end{figure}

We compare with our global MHD simulation to verify the picture of the weakly forward or reverse polar cap convection. Before $\Delta P_{dyn,1}$ impacted the simulated magnetosphere, reverse sunward convection was observed in the simulated polar ionosphere (not pictured), indicative of open-field-line reconnection poleward of the cusps. This is consistent with expectations based on the measured PCN index (Fig \ref{mhdpc}a). Very weak $<$0.5 km/s dawn-dusk asymmetric noon-to-midnight convection across the polar cap is observed in the simulated northern hemisphere shortly after the $\Delta P_{dyn,1}$ impact, indicative of low-latitude closed-field-line reconnection (Fig. \ref{mhdpc}b-c). Much weaker flows across the northern polar cap are observed later in the simulation, prior to the impact of the second pressure pulse (Fig. \ref{mhdpc}f-g). Convection in the southern hemisphere is mostly observed in a single-cell pattern around the polar cap boundary (Fig. \ref{mhdpc}d-e and \ref{mhdpc}h-i), particularly in the latter portion of the preconditioning phase, indicating the predominance of open-field-line magnetopause reconnection for a strong IMF $B_Y<0$\cite{ReiffandBurch.1985}. The lowest latitude of the simulated northern polar cap boundary expanded equator-ward by over $5^\circ$ between 4:00 and 5:00 UT (thick black lines, Fig. \ref{mhdpc}b,d), which is quantitatively inconsistent with the $2^\circ$ equator-ward expansion determined from SuperDARN (pink dashed line, Fig. \ref{s}g). The equator-ward growth of the polar cap boundary indicates loading of opened field lines from the day-side to night-side.

In summary, the MHD model and data agree that weak convection occurred during the preconditioning phase, which was dominated by a single-cell-type motion driven by open-field-line reconnection. While the simulated flux transfer rate was weak, as is also evidenced by the data, the simulated rate was likely overestimated.

\subsection{Magnetotail dynamics during preconditioning}

Given the weak solar wind driving, weak or suppressed low-latitude reconnection rate, and the lack of evidence for significant day-to-nightside flux transport at high latitudes, it is perhaps surprising to see the dramatic transformation of the magnetotail current sheet that took place during the preconditioning phase (Fig. \ref{s}j-p). This juxtaposition motivated the layout of Fig. \ref{s}; the stretching (Fig. \ref{s}k), thinning (Fig. \ref{s}l), and intensification (Fig. \ref{s}m) of the cross-tail current sheet are observed during the first dynamic pressure pulse and as the dynamic pressure subsided, in the absence of any discernible external driver. The thinning of the current sheet is also visible in the growth of ion thermal pressure non-gyrotropy (Fig. \ref{s}o), which grew to a few percent prior to the arrival of $\Delta P_{dyn,2}$, indicating that the current sheet was thin enough to support meandering ion motions \cite{ZenitaniandNagai.2016}. Figures \ref{s}a and \ref{s}n show that the magnetotail and solar wind pressures evolved in lock step, both increasing by +55\% during the passage of $\Delta P_{dyn,1}$, then both decreasing back to their initial values. We interpret this as evidence that the potentially weak flux loading had not substantially altered the magnetopause flaring angle. If the flaring angle had changed significantly, then the tail pressure at the end of preconditioning would have been significantly different from its initial value, given that the initial and final solar wind pressures were very similar. 

Still, and regardless of whether the magnetotail was or was not weakly driven by flux loading, we seek to understand how the tail current sheet thickness was reduced by roughly a factor of 5 and the northward equatorial magnetic field was reduced by over 2 orders of magnitude. The thickness of a Harris current sheet is controlled by the balance of the internal thermal and the external magnetic pressures. From 3:10 to 3:52 UT, the current sheet thinning may be explained by external compression resulting from $\Delta P_{dyn,1}>0$. What is needed is an explanation for the thinning between 3:52 to 5:14 UT, which occurred while the tail and solar wind pressures decreased. 

\begin{figure}
\noindent\includegraphics[width=39pc]{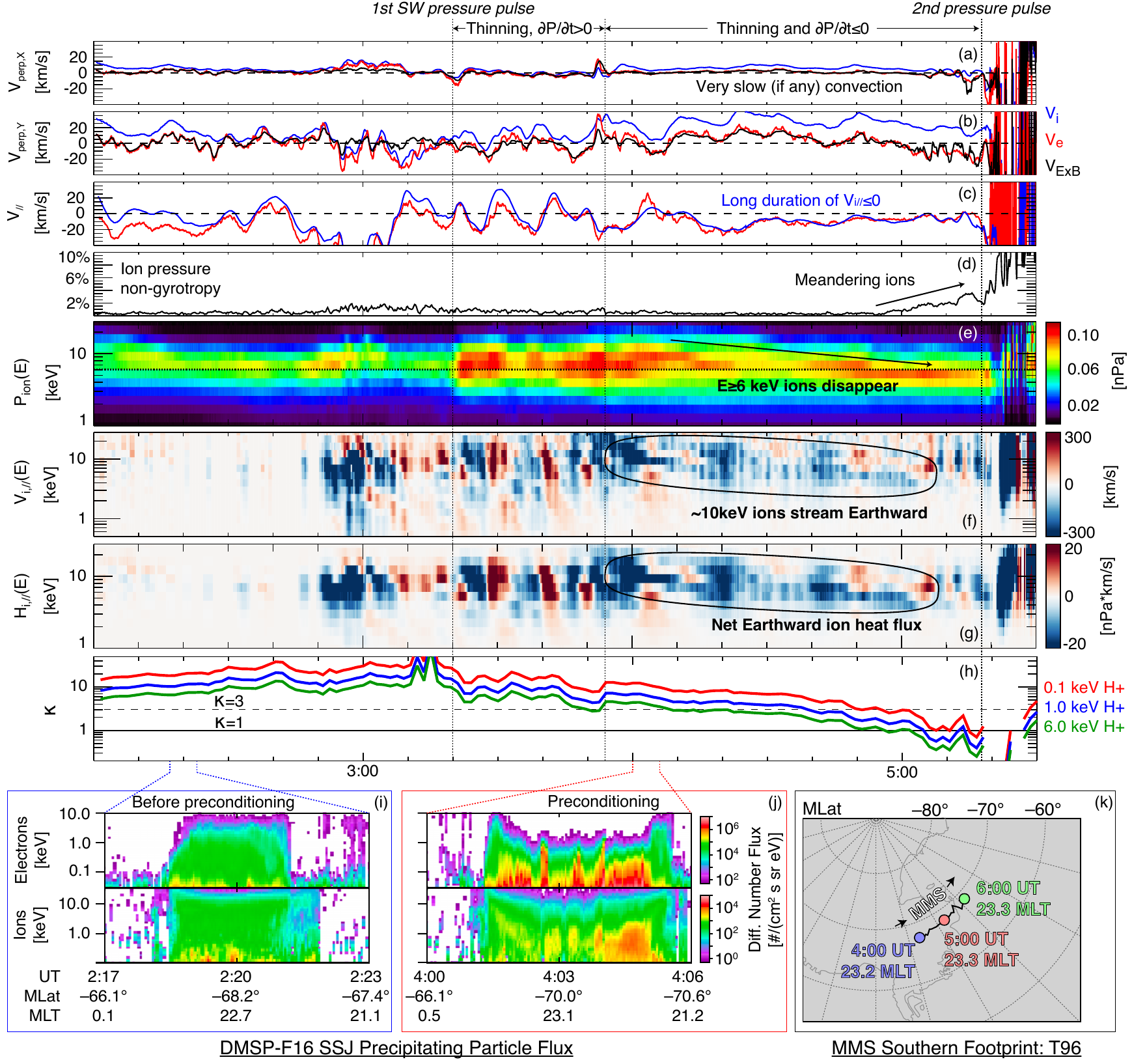}
\caption{(a)-(b): the $X_{GSM}$ and $Y_{GSM}$ components of the perpendicular (blue) ion, (red) electron, and $E\times B$ velocities. (c): the parallel velocities, with the same color code as (a) and (b). (d): ion non-gyrotropy from the $\sqrt{Q}$ parameter. (e), (f), and (g): the partial-energy ion scalar pressure (i.e., scalar pressure per energy bin), parallel ion velocity, and parallel ion heat flux respectively. (h): the ion adiabaticity parameter $\kappa$ for three energies (0.1 keV, 1 keV, 6 keV). (i)-(j): DMSP-F16 ion and electron flux data near the MMS foot point (i) before and (j) during the preconditioning. (k) The location of MMS mapped along empirically-modeled (using T96) magnetospheric field lines to their footpoints in the southern ionosphere.}
\label{precond}
\end{figure}

In the absence of external compression, the thickness of a one-dimensional current sheet can only be reduced by depleting the equatorial thermal pressure. The magnetic flux depletion mechanism of \citeA{HsiehandOtto.2015} drives thinning by depleting high-entropy flux tubes from the central plasma sheet. Magnetic flux depletion is manifested as the slow sunward convection of flux tubes, which are then convected in the azimuthal direction near the the plasma sheet-dipole boundary region \cite{Sun.2017}. This convection, however, is driven by pressure gradients established when the magnetopause is significantly eroded via low-latitude dayside reconnection, which is not observed for this event. This flux depletion mechanism may only be relevant in the near-Earth plasma sheet, also. Further evidence that this mechanism is not relevant for this event's preconditioning phase is that there is no discernible sunward convection at MMS (Fig. \ref{precond}a).

We also consider the possibility that current sheet was being torqued by high-latitude lobe reconnection, which allows the IMF $B_Y$ to penetrate into the magnetosphere. During the preconditioning, MMS observed consistently dusk-ward $B_Y$, which is directed oppositely to the dawn-ward IMF $B_Y$. Thus, we conclude that the high-latitude reconnection did not significantly impact the tail geometry at the MMS location. 

Non-adiabatic particle meandering can move particles – particularly high-energy ions, which contribute to the internal current sheet thermal pressure – across flux tubes. Duskward ion velocities of 20-to-30 km/s, faster than the $\vec{E}\times\vec{B}$-drift velocity, are observed by MMS during the preconditioning (Fig. \ref{precond}b). These duskward-moving ions are not correlated with significant pressure non-gyrotropies, however, indicating that they are likely diamagnetic motions, which do not drive particle transport. Thus meandering-driven thermal pressure loss does not likely explain the current sheet thinning.

Lastly, we note the presence of small but persistent field-aligned ion and electron bulk velocities between $-20\lesssim V_{\parallel} \lesssim 0$ km/s (Fig. \ref{precond}c). These very small bulk velocities are carried predominantly by fast-moving thermal ions with energies greater than $\gtrsim$6 keV (Fig. \ref{precond}f), and their earthward motion out of the plasma sheet is visible as as a moderate heat flux ($0\gtrsim H_{i\parallel}\gtrsim-20$ nPa*km/s; Fig. \ref{precond}g). The departure of the $\gtrsim$6 keV ions may constitute a substantial net loss of thermal energy from the plasma sheet as, early in the preconditioning phase, these high-energy ions contributed significantly to the bulk thermal pressure (Fig. \ref{precond}e). The non-zero heat flux also indicates a violation of entropy conservation in the plasma sheet at MMS (though strictly speaking this requires a non-zero $\nabla\cdot\vec{H}$). Non-adiabatic evolution of the current sheet and the breakdown of entropy conservation is understood to be a critical step toward enabling instability growth, e.g., ballooning and flux transport \cite{Birn.2009}. If these ions are truly lost from the plasma sheet, for instance if they have been pitch-angle-scattered into the loss cone and are lost to the neutral atmosphere, then the plasma sheet volume must decrease (i.e., thinning) to preserve pressure equilibrium with the lobes and shocked solar wind. The analogy, again, is like letting air out of a balloon to deflate it (or, slightly more accurately but less common in practice, putting the balloon in a refrigerator and letting the atmosphere compress it).

Figure \ref{precond}h-k investigates whether the ions are indeed pitch-angle scattered into the loss cone by (1) examining whether the plasma sheet conditions favor scattering and (2) examining particle fluxes in the ionosphere at the MMS foot point. Typically, the threshold $\kappa\leq3$ (see introduction) is used to identify when pitch-angle scattering is enabled. The curvature parameter $\kappa$ is calculated for 0.1 keV, 1 keV, and 6 keV protons in Fig. \ref{precond}h. The current sheet thickness was too large during this stage of preconditioning to be resolved by the electron-scale MMS tetrahedron, so $R_c$ is calculated as $R_c\approx hB_z/B_0$, as in \cite{BuchnerandZelenyi.1989} equation 4''. As is demonstrated by Fig. \ref{precond}h, $\kappa\lesssim3$ was reached for 6 keV protons during the initial solar-wind-driven compression (interval between the first two vertical lines in Fig. \ref{precond}h). $\kappa$ continued decreasing as the solar wind dynamic pressure abated (between the second and third vertical dashed lines). As $\kappa$ became smaller than 1, i.e., the curvature radius became much smaller than the gyroradius, MMS observed a steady increase in the ion pressure non-gyrotropy, which is consistent with ordered non-adiabatic meandering ion motions, as expected for $\kappa<<1$. As the plasma sheet $\kappa$ transitioned from the pitch-angle scattering regime to the meandering orbit regime MMS observed a reduction or cessation in the net Earthward flow of ions (Fig. \ref{precond}c and \ref{precond}f). 

DMSP F-16 crossed the ionosphere at high southern latitudes near the MMS footpoint (Fig. \ref{precond}k) twice during the interval of Fig. \ref{precond}. Comparing the crossings before (Fig. \ref{precond}i) and during preconditioning (Fig. \ref{precond}j) indicate that downward fluxes of ions with plasma sheet-like energies ($\sim$hundreds eV to several keV) increased by up to two orders of magnitude during preconditioning. 

In summary, we conclude that (1) increased pitch-angle scattering removed high-energy thermal protons from the central plasma sheet by scattering them into the loss cone, and (2) this ``lost'' population had contributed significantly to the plasma sheet thermal pressure at the start of the preconditioning phase. It is possible that this scattering-driven proton loss contributed to (rather than resulted from) current sheet thinning, with a simple analogy being drawn to letting the air out of a balloon (though in our case, the loss of thermal energy rather than number density is likely more relevant). More theoretical work is needed to evaluate whether this precipitative loss of particles can substantially thin the plasma sheet. While scattering-driven losses of hot ions likely contribute to the correlation between current sheet stretching ($\partial B_Z/\partial t<0$) and plasma sheet cooling (see \citeA{Runov.2021} and references therein); however, the temperature decrease is typically associated with the influx of cold ions from the ionosphere and flanks \cite{Artemyev.2019}. Note, however: given that flux loading at high latitudes and flux tube depletion at low latitudes are not operating in any discernible capacity, it is not clear how else the plasma sheet may have become thin during the preconditioning phase. 

\subsection{Impact of second solar wind pressure pulse and reconnection}

By 5:08 UT, the cross-tail current sheet had been thinned substantially and the north-south component of the plasma sheet magnetic field had been reduced nearly to zero; i.e., the conditions that had given stability to the magnetotail current sheet had been eroded. At 5:08 UT, ACE detected a second smaller transient pulse in the solar wind dynamic pressure (second vertical dashed line, Fig. \ref{s}a). Upon the arrival of the pulse, the magnetotail current sheet evolved rapidly. The half-thickness collapsed below the ion inertial scale (roughly 0.1-to-0.2 $R_E$ at 5:08 UT). Low and then high-frequency flapping-mode waves were observed (Fig. \ref{s}j). The north-south component of the plasma sheet magnetic field ceased its monotonic approach toward 0 nT and became highly structured, with strong positive and negative values being observed shortly after the current sheet collapse began (Fig. \ref{s}k). The magnetotail total pressure – rather than rising with the solar wind dynamic pressure – began a precipitous fall (Fig. \ref{s}n). MMS also observed an explosive growth of the ion non-gyrotropy (Fig. \ref{s}o). Lastly, near 5:10 UT, MMS observed the growth and then reversal of a fast ion jet (Fig. \ref{s}p), characteristic of reconnection.

The importance of the preconditioning is evidenced by the fact that the first larger solar wind pressure pulse did not trigger reconnection while the second pulse did. Based on the nearly simultaneous timing of the solar wind pressure pulse arrival, the current sheet collapse, and the reconnection onset, we conclude that the deformation of the current sheet boundary conditions by the deformation of the high-latitude magnetotail by the solar wind pressure pulse precipitated the loss of current sheet equilibrium and reconnection onset, as has previously been suggested \cite{BirnandSchindler.2002,Birn.2004}. The companion study, Paper 2, investigates the interval near reconnection onset in greater detail.

The reconnection was ultimately short lived, with the X-line being ejected tail-ward immediately after onset, and the recovery of the current sheet thickness began nearly immediately after reconnection onset (Fig. \ref{s}l). The reconnection also had a negligible impact on the inner magnetosphere. Geosynchronous satellites did not observe any discernible increase in energetic particle fluxes in the minutes after the onset (not pictured). A very weak deflection of the Auroral electrojet (AE), a proxy for substorm activity, to -50 nT was the only discernible global signature of the reconnection (see Paper 2). The lack of global substorm activity is also evidenced by the absence of this 3 July 2017 5:20 UT event from commonly used substorm lists based on multiple disparate sets of criteria for defining a substorm \cite{Newell.2011,Forsyth.2015,OhtaniandGjerloev.2020}.

\section{Summary, conclusions and open questions}

We investigated the global solar wind-magnetospheric interaction that lead to the initiation of magnetotail reconnection during an otherwise quiet interval on July 3, 2017. ACE data from the upstream solar wind identified two transient solar wind pressure pulses as the likely drivers of MMS-observed magnetotail dynamics. THEMIS-D and E data demonstrated the lack of any significant low-latitude magnetopause reconnection signatures, with the possible exception being two $\sim$50 km/s plasma flows near 5:10 UT, both observed by THEMIS-D. THEMIS-E, which was lower in altitude than THEMIS-D, observed a significant compression of the magnetopause after the arrival of the first solar wind pressure pulse, which also lead to a 4-fold increase in the magnetosheath density at THEMIS-D. Based on the modeled and observed magnetopause shear angle and the observed magnetopause $\Delta\beta$, we concluded that magnetopause reconnection may have been suppressed at low latitudes. This conclusion was consistent with (1) AMPERE data, which showed no or very weak field-aligned currents at the low-latitude dayside magnetopause, and (2) the SuperDARN-derived cross-polar-cap potential and convection boundary location, both of which indicated that convection and flux loading were either not occurring or very weak. A global MHD simulation was investigated to verify the weakly-driven nature of the polar cap. 

MMS observations indicated that the first solar wind pressure pulse triggered the gradual thinning and stretching of the cross-tail current sheet, which continued as the solar wind pressure abated. We suggested a new mechanism to explain this thinning; namely, the evacuation of the plasma sheet thermal pressure by pitch-angle scattering and precipitation. Pitch-angle scattering and precipitation were both apparently active, as observed by MMS and DMSP F-16, respectively. The plausibility of this mechanism for thinning the current sheet was not evaluated from a quantitative and theoretical perspective. Regardless of the cause, we demonstrate that substantial deformation and destabilization of the cross-tail current sheet is possible with the absence of strong solar wind driving by southward IMF. Finally, we point out the importance of the preconditioning interval is demonstrated that the two identical solar wind drivers (two pressure pulses) elicited dramatically different responses before and after the current sheet thinning. 

There are many open questions raised by this work. Since we do not have access to auroral images given as this event took place near the northern summer solstice: (1) what, if any, are the auroral signatures of this type of quiescent magnetotail reconnection? (2) Is this a truly viable mechanism for current sheet thinning, and if so then does it play a role in more intense substorms? (3) Why did this reconnection event not trigger a substorm, and (4) why were there no energetic particle injections observed at geosynchronous orbit? Only a fraction of these questions may be answerable with the present heliophysics system observatory.

\acknowledgments
KJG, CJF, and RBT were supported by NASA’s MMS FIELDS contract NNG04EB99C. We acknowledge NASA contract NAS5-02099 and V. Angelopoulos for use of data from the THEMIS Mission. We acknowledge the contributions made by many to collect and distribute the datasets used in this study. 

\section{Open Research}
MMS data, including FGM \cite{MMS_FGM_BRST}, FPI \cite{MMS_DES_MOMS_BRST,MMS_DIS_MOMS_BRST}, and EIS \cite{MMS_EIS_EXTOF_BRST,MMS_EIS_PHXTOF_BRST}, were obtained from https://lasp.colorado.edu/mms/sdc/public. DMSP data were obtained from https://dmsp.bc.edu. THEMIS ESA \cite{McFadden.2008} and FGM \cite{Auster.2008} data were obtained from http://themis.ssl.berkeley.edu. AMPERE data \cite{Waters.2020} were obtained from https://ampere.jhuapl.edu/. SuperDARN data were obtained from http://vt.superdarn.org. ACE data were obtained from  https://cdaweb.gsfc.nasa.gov. Substorm event lists are maintained by SuperMAG and are available at https://supermag.jhuapl.edu/substorms/. LANL SOPA data are not publicly available, and can be obtained for limited case studies by inquiry to the SOPA team (see https://cdaweb.gsfc.nasa.gov/Recent\_LANL\_Data.html). The Space Physics Environment Data Analysis System (SPEDAS) software package \cite{spedas} was used.


%
%


%
%
%
%
%

\end{document}